# CUDA BASED PERFORMANCE EVALUATION OF THE COMPUTATIONAL EFFICIENCY OF THE DCT IMAGE COMPRESSION TECHNIQUE ON BOTH THE CPU AND GPU

Kgotlaetsile Mathews Modieginyane[1], Zenzo Polite Ncube[2] and Naison Gasela[3]

School of Mathematics and Physical Sciences, North West University, Mafikeng Campus, Private Bag X2046, Mmabatho, 2735, South Africa
lekgotla.magivo@gmail.com
Zenzo.Ncube@nwu.ac.za
Naison.Gasela@nwu.ac.za

*ABSTRACT*

*Recent advances in computing such as the massively parallel GPUs (Graphical Processing Units),coupled with the need to store and deliver large quantities of digital data especially images, has brought a number of challenges for Computer Scientists, the research community and other stakeholders. These challenges, such as prohibitively large costs to manipulate the digital data amongst others, have been the focus of the research community in recent years and has led to the investigation of image compression techniques that can achieve excellent results. One such technique is the Discrete Cosine Transform, which helps separate an image into parts of differing frequencies and has the advantage of excellent energy-compaction.*

*This paper investigates the use of the Compute Unified Device Architecture (CUDA) programming model to implement the DCT based Cordic based Loeffler algorithm for efficient image compression. The computational efficiency is analyzed and evaluated under both the CPU and GPU. The PSNR (Peak Signal to Noise Ratio) is used to evaluate image reconstruction quality in this paper. The results are presented and discussed.*

*KEYWORDS*

*Compute Unified Device Architecture, Peak Signal to Noise Ratio, Graphical Processing Units, Discrete Cosine Transform,    Computational Efficiency.*

## 1. INTRODUCTION

The massive amount of digital data on the web has brought a number of challenges such as prohibitively high costs in terms of storing and delivering the data. Image compression therefore involves reducing the amount of memory it takes to store an image in order to reduce these large costs.

Image processing generally, is a very compute intensive task. Taking into consideration the image representation and quality, systems or application/techniques for image processing must have special capabilities for unquestionable results in their image processing. The rapid growth of digital imaging applications, including desktop publishing, multimedia, teleconferencing, computer graphics and visualization and high definition television (HDTV) has increased the need for effective image compression techniques.

Many image processing algorithms require dozens of floating point computations per pixel, which can result in slow runtime even for the fastest of CPUs. Because of the need for high performance computing in today's computing era, this paper presents a study based on Cordic based Loeffler DCT for image compression using CUDA. Image processing takes the advantage of CUDA processing because of the parallelism that pixels exhibit in an image and that can be offered by the CUDA architecture. In that way the CUDA architecture is found to be relevant for image processing [1], [2], [3] and [4].

This paper presents a performance based evaluation of the DCT image compression technique on both the CPU and GPU using CUDA. The focus of the paper is on the Cordic based Loeffler DCT.

We begin by giving a brief review of literature (background), followed by a description of the research methodology employed, including an overview of image compression, GPUs, CUDA architecture, Cordic based Loeffler DCT and related theory. This is then followed by the results of the experiments and their interpretation and lastly, conclusions and future work.

## 2. RELATED WORK

### 2.1 Computational Efficiency

Generally, computational efficiency refers to the proficiency of a machine or properties of an algorithm relating to the ratio of actual operating time to the scheduled operating time.

It can also be explained using an aspect of throughput, which is the amount of work that a computer can do over a period of time. Another measure of computer efficiency is performance, that is, the speed with which one or a set of batch programs run with a certain workload or how many interactive user requests are being handled with what responsiveness. A function or algorithm is computationally efficient if it is fast in its processing, it requires less memory and storage space for its operation, and consumes less power in its computation.

In the context of image compression including the above mentioned properties and conditions for proficient processing, computational efficiency of a transform code or method has other important aspects to be taken into consideration such as; image representation and image quality. A transform method is computationally efficient in its implementation for image compression if it has the capabilities of the above mentioned properties and does not compromise the representation and quality of the original input image.

### 2.2 Image compression

Image compression techniques are either Lossy or Lossless. By lossless and lossy compression we mean whether or not, in the compression of a file, all original data can be recovered when the file is uncompressed. With lossless compression, every single bit of data that was originally in the file remains after the file is uncompressed. All of the information is completely restored. This is a relevant technique for text or spread sheet files, where losing words or financial data could result in a very serious problem. An example of lossless image compression format is the Graphics Interchange File (GIF) which is normally used to represent web images.

Image compression is minimizing the size in bytes of a graphics file without degrading the quality of the image to an acceptable level. The primary objectives for image compression are to allow us to store image data in a given amount of storage space, as image data requires a lot of storage space, and to reduce the time required to transfer image data on the internet. It addresses the problem of image data storage by reducing the amount of data required to represent a digital image. This is achieved by removing data redundancy while preserving information content.
Image Compression addresses the problem of reducing the amount of data required to represent the digital image without losing its quality. Compression is achieved by the removal of one or more of three basic data redundancies: (1) Coding redundancy, which is present when less than optimal (i.e. the smallest length) code words are used; (2) Interpixel redundancy, which results from correlations between the pixels of an image; and (3) psycho visual redundancy which is due to data that is ignored by the human visual system (i.e. visually nonessential information).A detailed analysis of various image compression methods is found in [1], [2],[3],[4],[5] and [6].

**2.3 GPU (Graphics Processing Unit)**

A GPU (Graphics Processing Unit) is NVIDIAs core for graphics processing developed by NVIDIA in 1999. It is a specialized circuit designed to accelerate the image output in a frame buffer intended for output to a display. According to [7], the first GPU developed was a GeForce 256. This GPU model could process 10 million polygons per second and had more than 22 million transistors. GPUs are very efficient at manipulating computer graphics and are generally more effective than general-purpose CPUs for algorithms where processing of large blocks of data is done in parallel.

GPUs are highly parallel programmable cores on NVIDIA developmental platforms, and offer great relevance for high performance computing. In this research, the GPU hardware that will be used is the GeForce GTX 480, which is based on the Fermi architecture. This GPU architecture comprises of 480 cores optimized, using the Fermi architecture for efficient processing. GPUs have parallel throughput architecture that emphasizes many concurrent threads. This distinction makes them suitable for parallel programming. [7], [8] and [9].

Recently GPU processors are integrated onto CPU host computer systems. These computer systems are called heterogeneous systems. That is, a computer system consisting of a CPU as a host and a GPU as a device. In this architecture the CPU is the global execution controller (access the whole global memory) and the GPU uses local memory to perform computations. GPUs have parallel throughput architecture that emphasizes many concurrent threads. Consequently, parallel portions of applications called kernels are executed on the GPU (Device).

## 2.4 CUDA Architecture

CUDA (Compute Unified Device Architecture) developed by NVIDIA, is a parallel computing platform and programming model that enables efficient computing performance by coupling the power of the graphics processing unit so that it is a scalable parallel programming model and a software environment for parallel computing.

The CUDA architecture provides developers with a way to efficiently program GPUs using minimal extensions to C/C++ environments and it is a heterogeneous serial-parallel programming model. CUDA programming is based on the data parallel processing model and exhibits great relevance for compute intensive tasks such as image processing, computer visualization, scientific computing, etc. CUDA enables this outstanding performance through its standard APIs such as OpenCL and DirectX Compute, high level languages such as C/C++, FORTRAN, Java, Python and the Microsoft .NET framework as highlighted in [7] and [8].

According to [7], CUDA exposes a fast memory region that can be shared amongst threads and allows threads in the same block to coordinate their activities using a barrier synchronization function. The CUDA shared memory architecture makes it possible for thread cooperation.

## 2.5 DCT and the DCT Algorithms

The notation and definitions about the DCT algorithm in this paper were adopted from [10].

### 2.5.1 The DCT (Discrete Cosine Transform)

For a clear understanding of the DCT transform coding, particularly the 2-D DCT often called type-II DCT, a clear definition of the 1-D DCT will first be given towards the 2-D DCT definition itself.

**The 1-D DCT**

The equation below is a formal definition of the 1-D DCT, defined within 1-D sequence of length N points. *F(x)* is the 1-D DCT of a signal *f(i)* and $\propto (i)$ is for normalizing the basis function which is the cosine term.

$$F(x) = \left(\frac{2}{N}\right)^{\frac{1}{2}} \sum_{i=0}^{N-1} \alpha(i) . \cos\left[\frac{\pi . u}{2 . N}(2i+1)\right] . f(i) \qquad (3)$$

and the corresponding inverse 1D DCT transform is given as $F^{-1}$ that is: where

$$\alpha(i) = \begin{Bmatrix} \frac{1}{\sqrt{2}} & \varepsilon = 0 \\ 1 & otherwise \end{Bmatrix}$$

(4)

**The 2-D DCT**

The general equation (for an NxM image) of the 2-D DCT is defined by the following equation (6), where *F(u, v)* is the 2-D DCT of a 2-D DCT signal *f(i, j)*.

$$F(u,v) = \left(\frac{2}{N}\right)^{\frac{1}{2}} \left(\frac{2}{M}\right)^{\frac{1}{2}} \sum_{i=0}^{N-1} \sum_{J=0}^{M-1} \alpha(i).\alpha(j) \left[\frac{\pi.u}{2.N}(2i+1)\right] \cos\left[\frac{\pi.v}{2.M}(2j+1)\right].f(i,j) \quad (6)$$

with its corresponding inverse (2D IDCT) transform given as $F^{-1}(u,v)$, that is:
where:

$$\alpha(\varepsilon) = \begin{cases} \frac{1}{\sqrt{2}} & \varepsilon = 0 \\ 1 & otherwise \end{cases} \quad (7)$$

### 2.5.2 The DCT Algorithms

For a better understanding of the Cordic based Loeffler DCT, we start with a brief description of the Loeffler DCT algorithm itself. The Loeffler algorithm has a combination of four stages. All these stages (1 - 4), need to be executed in serial mode because of the data dependencies involved during processing. Prallelization, though, can still be achieved in calculations inside one stage. In stage 2, the algorithm splits in two: one part for the even coefficients and the other for the odd ones. The even part is actually a 4-points DCT, which is then separated in even and odd parts in stage 3. The Cordic based Loeffler DCT, which is a derivation from the Loeffler DCT is made by rotating all the input values by $\frac{\pi}{4}$ using the rotation block and its associated equations (which is atage 4). Figure 1 below illustrates the Cordic based Loeffler DCT flow graph.

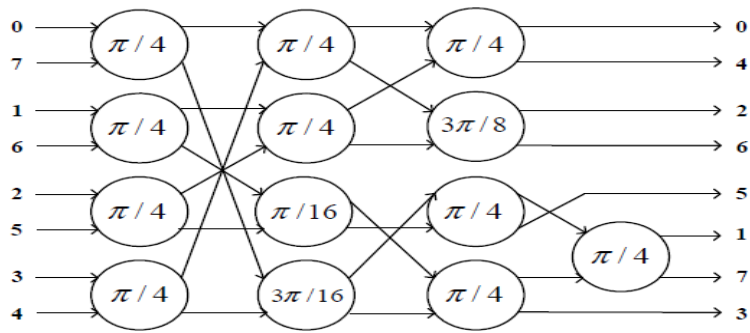

Figure 1. Flow graph of the Cordic based Loeffler DCT (Source:[11])

Researchers such as [13] and [14] explored the relationship between image compression and the discrete cosine transform. These researchers analysed image compression using discrete cosine transform and highlighted its power in image processing. [14] went on to developed functions to compute the DCT to compress images.

Some researchers have also carried out an experimental evaluation of the DCT on the GPU and the CPU. [12] presented several techniques for efficient implementation of DCT on both the CPU and GPU using direct matrix multiplication. Using five methods for performing the DCT on the GPU a conclusion was made from the experimental analysis that the speed of the DCT on the GPU exceeded that on the CPU illustrating the efficiency of the DCT implementation on the GPU.

Castaño-Díez et al. in [15] used the CUDA platform to evaluate the performance evaluation of image processing algorithms on the GPU. Algorithms considered in their paper were spatial transformations, real-space and the Fourier operations. Its CUDA implementation revealed that, compared to the CPU implementations, the GPU implementations of the said algorithms were 10 -20 times faster, illustrating the power of the GPU in terms of processing computationally intensive operations as indicated in [10] and [16].

For a detailed analysis of the DCT algorithm and its variants, see [13],[17],[18] and [19].

## 3. RESEARCH METHODOLOGY

This research is based on an experimental performance evaluation of the selected DCT algorithm in a heterogeneous system comprising of a CPU and a GPU. In a heterogeneous system the CPU is called the host and the GPU is called the device. The CPU takes control of the system. Instructions to the device are copied from the host for processing and copied back to the host after being processed (by the device) for output or other serial workloads, using relevant device programmable model functions.

### 3.1 Experiment Settings

1.CPU:Intel 64 Bit Processor i3-2130
2. Memory:4GB
3.GPU:NVIDIA Fermi GeForce GTX 480, CUDA version 5.0
4.System :Windows 7

### 3.2 Comparison parameters

To examine computational efficiency of the DCT ,the authorr will use processing speed measured in milliseconds and the image size as throughput on both the CPU and the GPU.

The Cordic based Loeffler DCT algorithm will be implemented as a core for image compression using CUDA. This implementation will be done on several images with concrete analysis based on the image quality, compression loss, image errors, and image representation. CUDA C++ programming language will be used for all computations in this research. CUDA developmental tools namely; CUDA SDK and CUDA graphics drivers will be installed on the system for software and hardware compatibility support and usage of the GPU.

Images will be loaded from their directory to the sample for processing, and then assessed for compression by the working system, so as to launch the DCT implementation functions. The DCT, the quantizer and the IDCT executes on different kernels. Compressed images will be examined and the computational performance of the two cores and efficiency of the DCT will be critically analysed.

## 4. INTERPRETATION OF FINDINGS

In this research paper, the image quality after compression, DCT compression time for both CPU and GPU were studied and the speedup graph was also illustrated from the results, all contributing to the examination of the computational efficiency of the DCT.

The picture below (Figure 2) is a grayscale representation of Lena, on which the DCT transform will be computed on the CPU using serial code and on the GPU using parallel code. All original images used in this paper were taken from Marco Schmidt's standard database for test images used in image processing.

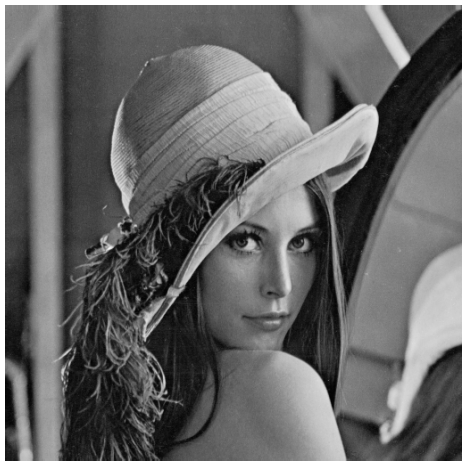

Figure 2. Lena original (gray).

The two pictures below, figures 3 and 4 are the CPU processed and the GPU processed images respectively using a 2048x2048 pixel intensity of Lena's image.

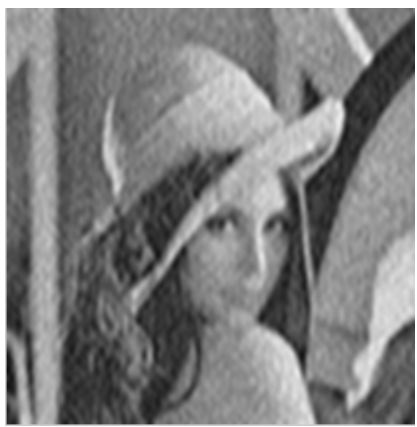
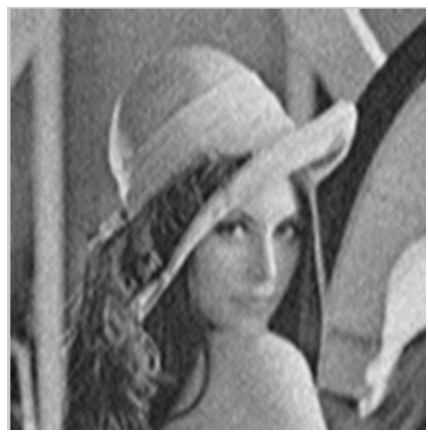

Figure 3. Lena CPU processed (serial code).　　　　　Figure 4. Lena GPU processed (parallel code).

The image quality processed using the CPU (Figure 3), exhibits a lot of degradation compared to the one processed using the GPU (Figure 4).

Table 1 below shows the time comparison of grayscale equalization about the CPU and the GPU on different sizes of Lena's image. Based on transformations which were done on different image sizes of Lena, it was found that the GPU speedup was significantly larger compared to the CPU.

Table 1: Time comparisons of grayscale histogram of Lena for both CPU and GPU.

| Input image | CPU(ms) | GPU(ms) |
|---|---|---|
| 3072x3072 | 1020.32 | 8.92 |
| 2048x2048 | 266.23 | 5.61 |
| 1600x1400 | 116.12 | 2.20 |
| 1024x814 | 88.23 | 1.24 |
| 576x720 | 48.52 | 0.82 |
| 512x512 | 16.42 | 0.62 |
| 200x200 | 6.88 | 0.24 |

The table shows that as the image size increases the CPU takes longer to process the image than the GPU does for the same picture size. These results are also displayed in Figures 5 and 6 below.

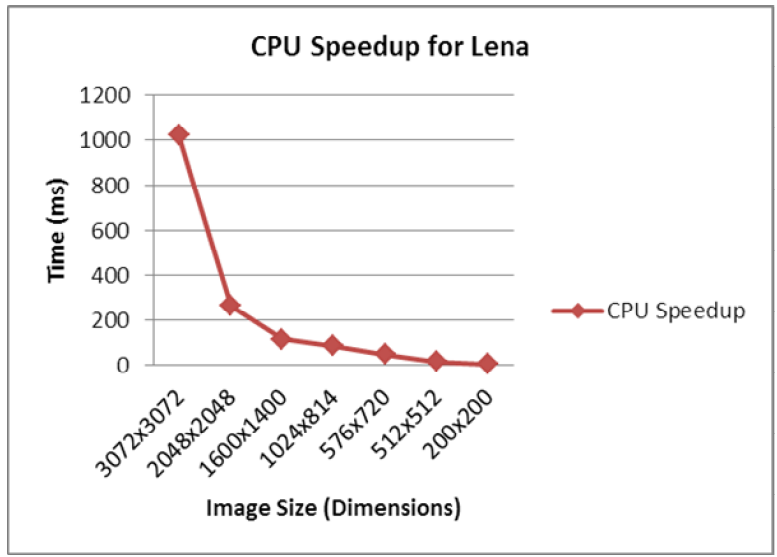

Figure 5. Speedup graph of Lena by the CPU.

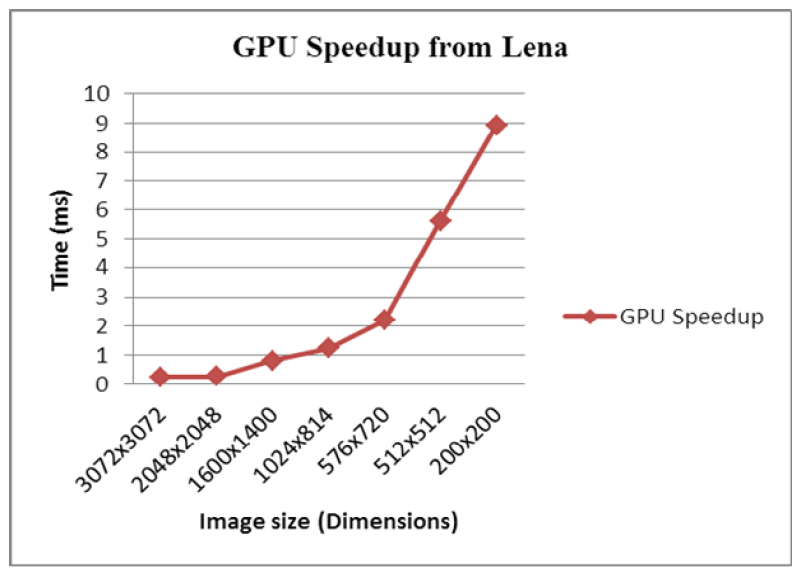

Figure 6. Speedup graph of Lena by the GPU.

Further experiments similar to those done on the Lena image above(on the CPU and GPU platforms) were carried out on a different picture (Figure 7).The picture below (Figure 7) is a grayscale representation of Cable-car, which was used in the second experiment.

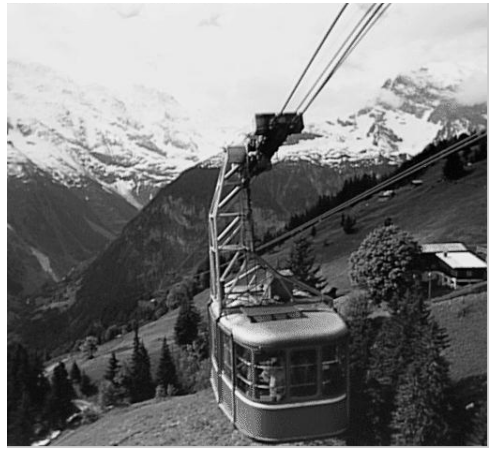

Figure 7. Cable-car original (gray).

The two pictures below, Figures 8 and 9 are the CPU processed and the GPU processed images respectively using a 544x512 pixel intensity of the Cable-car's image.

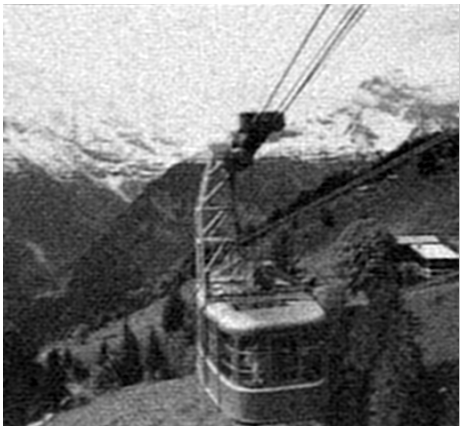 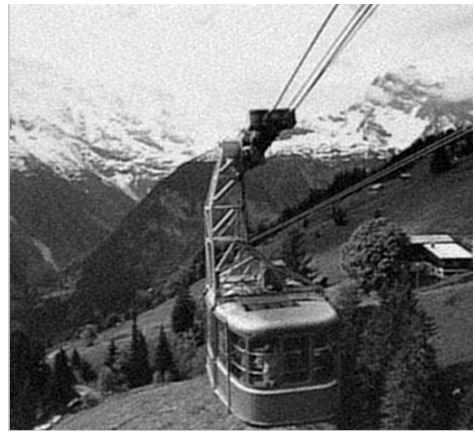

Figure 8. Cable-car CPU (serial code).              Figure 9. Cable-car GPU (parallel code).

The quality of the image processed using the CPU (Figure 8), exhibits a lot of degradation compared to the one processed using the GPU (Figure 9).

Table 2 below shows the time comparison of grayscale equalization about the CPU and the GPU on different sizes of the Cable-car image. Based on transformations which were done on different image sizes of Cable-car, it was found that the GPU speedup was significantly larger compared to the CPU.

Table 2. Time comparisons of grayscale of Cable-car for both CPU and GPU.

| Input image | CPU(ms) | GPU(ms) |
|---|---|---|
| 544x512 | 30.32 | 0.58 |
| 512x480 | 26.84 | 0.41 |
| 448x416 | 21.22 | 0.34 |
| 384x352 | 17.28 | 0.26 |
| 320x288 | 10.86 | 0.19 |

Table 2 shows that as the image size increases the CPU takes longer to process the image than the GPU does for same picture size. These results are also displayed in Figures 10 and 11 below.

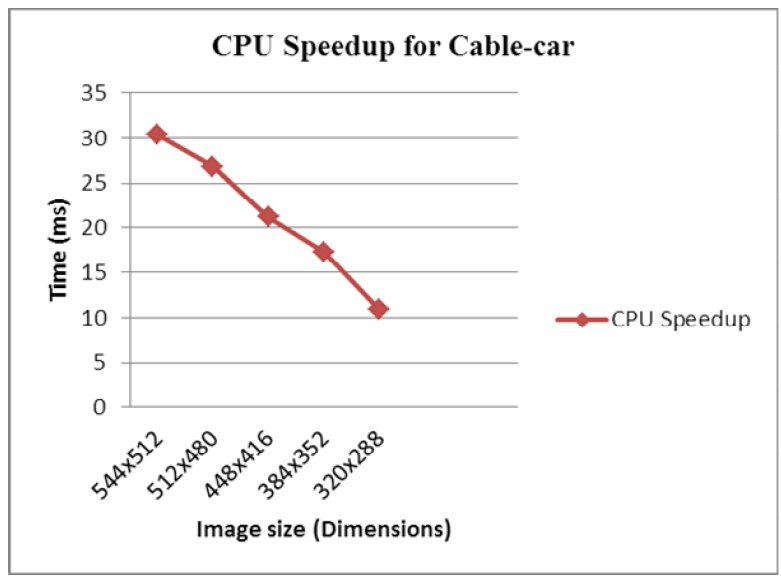

Figure 10. Speedup graph of Cable-car's by the GPU.

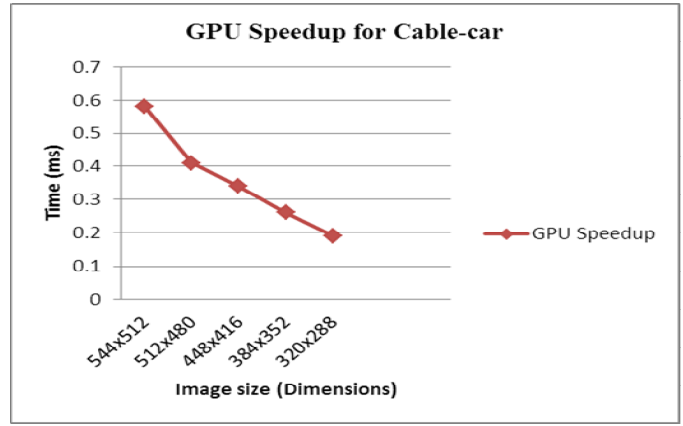

Figure 11. Speedup graph of Cable-car by the GPU.

## 4.1 PSNR (Peak Signal to Noise Ratio) Evaluation

To further examine the efficiency of the DCT, the PSNR comparison evaluation of the original DCT algorithm and the proposed algorithm was done on both Lena and Cable-car images. The tables bellow (Table 3 and 4) illustrates the PSNR evaluations of the mentioned images. The PSNR is used to evaluate the image reconstruction quality and it is expressed in decibels (dB). The PSNR defined for two images (original and compressed) O and C of size NxM, is defined by the equation (23) below as:

$$PSNR_{(O,C)} = 20 \log_{10} \frac{MAX}{\sqrt{MSE_{(O,C)}}} \qquad (23)$$

Where O is the original image and C is the reconstructed or noisy image. MAX is the maximum pixel value in image O, and MSE is the mean square error between O and C, MSE is given as:

$$MSE_{(C,O)} = \frac{1}{N}\frac{1}{M}\sum_{i=0}^{N-1}\sum_{j=0}^{M-1}\|O(i,j) - C(i,j)\|^2 \qquad (24)$$

The PSNR between the original and compressed images was calculated as below. The results are shown for both the DCT and Cordic based Loeffler DCT. The two tables (Table 3 and 4) below illustrate these results for Lena and the Cable-car images respectively for their different sizes.

Table 3. Lena PSNR values calculated between the original and compressed images.

| Lenna | 200x200 | 512x512 | 2048x2048 | 3072x3072 |
|---|---|---|---|---|
| **DCT** | 31.612543 | 33.188042 | 35.521183 | 37.077885 |
| **Cordic based Loeffler DCT** | 29.445233 | 31.157837 | 33.224584 | 35.111256 |

Table 4. Cable-car PNSR values calculated between the original and compressed images.

| Cable-car | 320x288 | 384x352 | 448x416 | 512x480 | 544x512 |
|---|---|---|---|---|---|
| **DCT** | 24.224891 | 26.154872 | 28.112488 | 30.224133 | 32.254781 |
| **Cordic based Loeffler DCT** | 21.275488 | 24.556324 | 26.985411 | 28.128771 | 30.845126 |

Careful observations about the PSNR values of the original and the proposed algorithm for the two experimental images were made, and it was found that; the Cordic based Loeffler DCT tends to be a good variation of the DCT and can work well for optimisation purposes.

## 5. CONCLUSIONS

An implementation of a CUDA based DCT image compression using a Cordic based Loeffler algorithm was done and the computational efficiency was analyzed and evaluated under both the CPU and GPU. Evidence about the computational efficiency of the DCT for image compression was found was found to be good, especially for compute intensive tasks such as image processing using NVIDIA GPUs. The DCT algorithm itself, maps well with parallel architectures, and that makes it still the best technique for image or digital data processing.

## 6. FUTURE WORK

- Other image compression techniques can be explored under the CUDA platform.
- The evaluation of this algorithm can also be evaluated under different GPUs other than the GTX 480.
- The experimental analysis can also be done using other tools besides CUDA and the results compared with those of using CUDA.
- Photo density variations should be taken into account to see how they affect the results.

## ACKNOWLEDGEMENTS

I would like to thank my mother Jane Modieginyane, brothers and sisters, who have given me their unequivocal support throughout, as always, for which my mere expression of thanks likewise does not suffice.

I would also like to acknowledge the financial, academic and technical support of the North West University for all the equipment they have provided, and the Department of Computer Science for their support and assistance since the start of my postgraduate studies in 2012.

I am most grateful to NVIDIA for their donation with the GeForce GTX 480, their all documented guidance for parallel programming using the GPU and CUDA architectures, to the success of my research project.

**Authors**

**Kgotlaetsile Mathews Modieginyane**

**BSc Computer Science and Electronics** 2009, North West University, South Africa. **BSc Honors Computer Science** (2012), North West University, South Africa. Currently (2013) doing **MSc Computer Science**. Research interest: High Performance Computing, Cloud Computing, Digital Signal Processing.

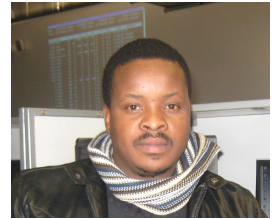

**Zenzo Polite Ncube**

**PhD** in Computer Science, Potchefstroom Campus of the North West University. He is currently employed as a Lecturer in the Department of Computer Science at the Mafikeng Campus of the North-West University, South Africa. Main areas: HPC, Cloud Computing, Networks.

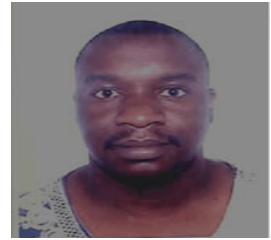

**Naison Gasela**

**PhD** in Computer Science at NUST Zimbabwe. Currently employed as a Lecturer in the Department of Computer Science at the Mafikeng Campus of the North-West University, South Africa. Main areas: HPC, Knowledge Discovery, Green Computing.

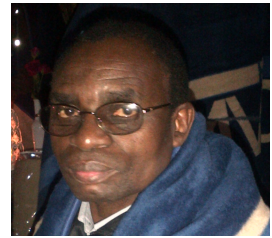